\documentclass[conference]{IEEEtran}
\usepackage{graphicx}

\usepackage{amsfonts}
\usepackage{amssymb,hhline}
\usepackage[ruled,boxed,boxruled]{algorithm2e}

\ifCLASSINFOpdf
\else
\fi
\usepackage[cmex10]{amsmath}
\usepackage{url}

\RequirePackage[citecolor=blue,urlcolor=blue]{hyperref} 

\graphicspath{{./}}
\DeclareGraphicsExtensions{.eps}

\newcommand{\paino}{\sl}

\newcommand{\bom}[1]{\boldsymbol{#1}}
\newcommand{\bo}[1]{\mathbf{#1}}

\newcommand{\s}{\bo x}  

\renewcommand{\a}{\bo a}  
\newcommand{\es}{x}      
\newcommand{\y}{\bo y} 
 
\newcommand{\ee}{r}      

\newcommand{\mm}{\bom \Phi}   
\renewcommand{\S}{\bo X}  
\newcommand{\Y}{\bo Y}  
\newcommand{\X}{\bo X}  
\newcommand{\E}{\bo R} 
\newcommand{\G}{\bo G} 
\newcommand{\A}{\bo A} 
\newcommand{\B}{\bo B} 
\newcommand{\SNR}{\mathrm{SNR}} 
\newcommand{\I}{\bo I} 

\newcommand{\W}{\bo W}

\newcommand{\Gam}{\Gamma} 
\newcommand{\sig}{\sigma}

\newcommand{\eps}{\bo e} 
\newcommand{\eeps}{e} 
\newcommand{\al}{\alpha} 
 
\newcommand{\bth}{\bom \theta} 

\newcommand{\be}{\alpha} 
\newcommand{\im}{\jmath} 
\newcommand{\ssgn}{\mathrm{sign}} 
\newcommand{\cov}{\mathrm{Cov}}

\newcommand{\Eps}{\bo E} 

\newcommand{\hop}{\mathrm{H}}

\newcommand{\R}{\mathbb{R}}    
\newcommand{\C}{\mathbb{C}}    

\newcommand{\supp}{\mathrm{supp}}
\newcommand{\rsupp}{\mathrm{supp}}

 \renewcommand{\vec}{\mathrm{vec}}
 
 \newcommand{\tr}{\mathrm{Tr}}
\newcommand{\expec}{\mathbb{E}}    

\newcommand{\pr}{\partial}

\newcommand{\beq}{\begin{equation}}
\newcommand{\eeq}{\end{equation}}
\newcommand{\bmat}{\begin{pmatrix}}
\newcommand{\emat}{\end{pmatrix}}
\newcommand{\beqa}{\begin{eqnarray}}
\newcommand{\eeqa}{\end{eqnarray}}

\newcommand{\mb}{\bom \phi}    
\newcommand{\ndim}{M}             
\newcommand{\pdim}{N}             
\newcommand{\kdim}{K}             
\newcommand{\qdim}{Q} 
\renewcommand{\Re}{\mathrm{Re}}

\newtheorem{definition}{Definition}

\begin{document}
%
\title{Multichannel sparse recovery of complex-valued signals using Huber's criterion }

\author{
\IEEEauthorblockN{Esa Ollila}
\IEEEauthorblockA{Department of Signal Processing and Acoustics, 
Aalto University \\ P.O.Box 13000,  FI-00076 Aalto, Finland} }
\maketitle

\begin{abstract}
In this paper, we generalize Huber's criterion 
to multichannel sparse recovery problem of complex-valued measurements where the objective is to find good recovery of jointly sparse
unknown signal vectors from the given multiple measurement vectors which are different linear combinations of the 
same known elementary vectors.  This requires careful characterization of robust complex-valued loss functions as well as 
Huber's criterion function for the multivariate sparse regression problem.  
We devise a greedy algorithm based on 
simultaneous normalized  iterative hard thresholding  (SNIHT) algorithm.  
Unlike the conventional SNIHT method, our algorithm, referred 
to as HUB-SNIHT, is robust under heavy-tailed non-Gaussian noise conditions, yet has a negligible performance loss compared to
SNIHT under Gaussian noise. 
Usefulness of the method is illustrated  in source localization application with sensor arrays. 
\end{abstract}


%
\IEEEpeerreviewmaketitle

\section{Introduction}

In the {\paino multiple measurement
vector (MMV) model},  a single measurement matrix is utilized
to obtain multiple measurement vectors, i.e., 
$\y_i = \mm \s_i + \eps_i,$  $i = 1,\ldots,\qdim$  
where 
$\mm  = \bmat \mb_{1} & \cdots & \mb_{\pdim}\emat = \bmat \mb_{(1)} & \cdots & \mb_{(\ndim)}\emat^\hop$ 
 is an $\ndim \times \pdim$ {\paino measurement matrix}  and $\eps_i$ 
are the (unobserved) random {\paino noise} vectors. Typically there are more column vectors   $\mb_{i}$ 
than row vectors $\mb_{(j)}$, i.e., $\ndim<\pdim$.  
 The unknown {\paino signal vectors} $\s_i$, $i=1,\ldots,\qdim$ are assumed to be {\paino sparse}, i.e., most of the elements are zero.  
 In matrix form, the MMV model is 
 \beq \label{eq:MMVmodel} 
\Y  = \mm \S + \Eps,
\eeq 
where $\Y = ( \y_1 \, \cdots \,  \y_{\qdim} ) \in \C^{\ndim \times \qdim}$, $\S=( \s_1 \ \cdots \ \s_{\qdim}) \in \C^{\pdim \times \qdim}$ and 
$\Eps =( \eps_1 \ \cdots \ \eps_{\qdim} ) \in \C^{\ndim \times \qdim} $ collect the measurement, the signal  and the error vectors, respectively. 
When $\qdim=1$, the model reduces to standard {\paino compressed sensing (CS) model} \cite{duarte_eldar:2011}. 
The key assumption of MMV model  
is that the signal matrix $\S$ is $\kdim$-rowsparse, i.e., at most  $\kdim$ rows of $\S$ contain
non-zero entries.  The {\paino row-support}  of  $\S$ is the index set of rows  containing non-zero  elements, 
$
\rsupp(\S) 
= \{   i \in \{1,\ldots,\pdim\} \, : \: \es_{ij} \neq 0 \,  \mbox{for some $j$} \}.
$
When $\S$ is $\kdim$-rowsparse, i.e., $| \rsupp(\S)| \leq \kdim$,   
joint estimation can lead  both to computational advantages and increased reconstruction accuracy; See
 \cite{tropp_etal:2006,tropp:2006,chen_huo:2006,eldar_rauhut:2010,duarte_eldar:2011,blanchard_etal:2014}. 

The objective of {\paino multichannel sparse recovery} 
problem is on finding a row sparse approximation of the signal matrix  $\S$ based on knowledge of $\Y$, the measurement matrix $\mm$ and
the sparsity level $\kdim$. 
Such a problems arises  
in electroencephalography and 
magnetoencephalography (EEG/MEG) \cite{duarte_eldar:2011} 
blind source separation \cite{gribonval_zibulevsky:2010}, 
and direction-of-arrival (DOA) estimation  of  sources in array and radar processing \cite{malioutov_etal:2005,wang_etal:2009,fortunati_etal:2014}. 
Many  greedy pursuit CS reconstruction algorithms 
have been extended for solving MMV problems. 
These methods, such as simultaneous normalized iterative hard thresholding (SNIHT) algorithm  \cite{blanchard_etal:2014} are guaranteed to perform very well provided that suitable conditions (e.g., incoherence  of $\mm$ and  non impulsive noise conditions) are met. The derived (worst case) recovery bounds depend linearly on $\| \Eps \|_2$, so the methods are not guaranteed to provide accurate reconstruction/approximation  under heavy-tailed non-Gaussian noise.  

In this paper, we generalize Huber's criterion \cite[cf. Section~7.7, 7.8]{huber:1981} (often referred to as "Huber's approach 2") originally developed for overdetermined  linear 
regression ($\ndim > \pdim$, $\qdim=1$) model to the {\it complex-valued} case and for the more general multivariate {\it sparse regression} problem.  This requires generalizing  robust $M$-estimates of 
regression (and loss  functions) for complex-valued  case.  In Huber's devise, one estimates the signal matrix and scale of the error terms simultaneously. This is necessary 
since most robust loss-functions require an estimate of the scale.  
Using Huber's criterion in the  MMV model one may elegantly estimate 
both the sparse signal matrix and the scale of the errors simultaneously. In particularly, we are able to circumvent the problem of obtaining a preliminary robust scale estimate 
which is a challenging problem in ill-posed multivariate sparse regression model since the support of $\S$ and hence the contributing elementary vectors of $\mm$ on measurements  
 are not known. 
In earlier related work Huber's approach 2 has been considered for Lasso-type real-valued linear regression setting in \cite{owen2007robust,lambert2011robust} 
and real-valued compressed sensing in \cite{ollila_etal:2014}. 
For our multichannel sparse recovery problem, we devise SNIHT algorithm 
which  results in a simple, computationally efficient 
and scalable approach for solving the MMV sparse reconstruction problem.

Let us offer a brief outline of the paper.  In Section~\ref{sec:backgr}, we give necessary notations and definitions as well as provide motivation and 
background of robust sparse recovery problem. 
Robust complex-valued loss functions and their properties  are outlined in Section~\ref{sec:loss} and a generalization of Huber's loss function for  complex measurements is given. Then, in Section~\ref{sec:hub} we formulate Huber's criterion for MMV model and the related SNIHT algorithm, called HUB-SNIHT, 
is derived in  Section~\ref{sec:SNIHT}. 
Finally, we illustrate the usefullness of the  method in  source localization application 
in Section~\ref{sec:array}.   
 
\section{Background} \label{sec:backgr}

\subsection{ Notations} 

For a matrix $\A \in \C^{\ndim \times \pdim}$  and an index set $\Gam$ of cardinality $|\Gam|=\kdim$, we denote 
 by   $\A_{\Gam}$  (resp. $\A_{(\Gam)}$) the $\ndim \times \kdim$ (resp. $\kdim \times \pdim$) matrix  
restricted to the columns (resp. rows) of $\A$ indexed by the set $\Gam$.
The $i$th column vector of $\A$ is denoted by $\a_i$ and the hermitian transpose of the  $i$th row vector of $\A$ by $\a_{(i)}$, 
$\A=(\a_1 \ \cdots \ \a_\pdim) = (\a_{(1)} \ \cdots \ \a_{(\ndim)} )^\hop$.  Furthermore, if $f: \C \to \C$, then $f (\A)$  refers to element-wise application of the function to its matrix valued argument, so $f(\A) \in \C^{\ndim \times \pdim}$ with $[f(\A)]_{ij}=f(a_{ij})$.

The usual Euclidean
norm on vectors will be written as $\| \cdot \|$. 
The matrix space $\C^{\ndim \times \pdim}$ is
equipped with the usual Hermitian inner product
\[
\langle \A, \B \rangle = \tr(\B^\hop \A) = 
\sum_{i=1}^{\ndim} \sum_{j=1}^{\pdim} a_{ij} b_{ij}^*
\]
where the trace of a (square) matrix is the sum of
diagonal entries. We define the weighted inner product as 
\[
\langle \A, \B \rangle_{\W} =   
\sum_{i=1}^{\ndim} \sum_{j=1}^{\pdim} w_{ij}a_{ij} b_{ij}^*
\]
where $\W$ is  $\ndim \times \pdim$ real matrix of positive weights. Note that $ \langle \A, \B \rangle_{\W}$ reduces to conventional inner product when $\W$ is a  matrix of ones. 
The Frobenius norm  is given by the inner product as
$
\| \A \|= \sqrt{\langle \A, \A \rangle} 
$
and $\| \A \|_{\W} =\sqrt{\langle \A, \A \rangle_{\W} } $ denotes the weighted Frobenius norm. 
The row-$\ell_0$ quasi-norm of $\A$ is the number of nonzero rows, i.e., 
$
\| \A \|_0 = | \ \rsupp(\A) | $.
 Hence the assumption that the signal matrix $\S \in \C^{\pdim \times \qdim}$ is   $\kdim$-rowsparse in the MMV model 
is equivalent with the statement that  $\| \S \|_0 \leq \kdim$.


We use  
$H_\kdim(\cdot)$ to denote the {\paino hard thresholding operator}: 
for a matrix $\S \in \C^{\pdim \times \qdim}$, 
 $H_\kdim(\S)$   retains the elements of the $\kdim$ rows of $\S$ that possess largest  $\ell_2$-norms and set elements of the other rows to zero.    
 Notation $\S|_{ \Gam}$ refers to sparsified version of $\S$ such that the entries
in the rows indexed by set $\Gam$ remain unchanged while all other
rows  have all entries set to $0$.

\subsection{Robust constrained optimization problem} \label{Ollila:sec2}

Suppose that the error terms $\eeps_{ij}$ are i.i.d. continuous random variables from a circular distribution \cite{ollila_etal:2011} with p.d.f. $f(e)=(1/\sigma)f_0(e/\sig)$, where 
$f_0(e)$ denotes the standard form of the density and $\sigma>0$ is the scale parameter.     
If the scale is known, then a reasonable approach for solving the simultaneous sparse recovery problem is 
to minimize a distance criterion of residuals,  
\beq \label{eq:D}
D_\rho\!\left( \frac{\Y- \mm \S}{\sig} \right)=\sum_{i=1}^\ndim \sum_{j=1}^\qdim  \rho\left( \frac{y_{ij}  - \mb_{(i)}^\hop \s_j}{\sig} \right)
\eeq 
for some suitable loss function $\rho(\cdot)$ 
subject to $\kdim$-rowsparsity constraint  $\| \S \|_0 \leq \kdim$. 
For conventional least squares (LS) loss function $\rho(e)=|e|^2$, the scale can be factored out from the objective function, and 
the minimization problem reduces to 
\[
\min_{\S} \| \Y - \mm \S \|^2 \quad \mbox{subject to} \quad \| \S \|_0 \leq \kdim. 
\]
The well-known problem with LS minimization is that it gives a very small weight on small residuals and
a strong weight on large residuals, implying that even a single large outlier can have a large influence
on the obtained result.  

At least two problems arises when using conventional robust loss functions in \eqref{eq:D}. 
First,  commonly used robust loss functions in robust statistics 
such as Huber's or Tukey's loss functions 
require an estimate of scale $\sig$. Obtaining a reliable robust estimate of scale is a difficult problem. It involves  obtaining a $\kdim$-rowsparse 
robust preliminary estimate $\hat \S_0$ of the signal matrix and then computing  robust scale estimate based on the resulting residual matrix $\E_0 = \Y- \mm \hat \S_0$. 
Second problem is that robust loss functions 
are defined in the real-valued case 
and some thought must be given  on special properties of  complex-valued loss functions. These problems are addressed next in Section~\ref{sec:loss} and 
Section~\ref{sec:hub}.

\section{Loss functions: complex valued case}  \label{sec:loss}

We start by giving a proper definition of 
a loss function $\rho$. 

\begin{definition}  Function $\rho: \C \to \R^+_0$ is called a {\paino loss function} if it verifies: 
\begin{itemize}
\item[{\bf (L1)}] $\rho$ is circularly symmetric,  $\rho(e^{\im \theta }x)=\rho(x)$, $\forall \theta \in \R$. 
\item[{\bf (L2)}] $\rho(0)=0$. Furthermore, $\rho$ is $\R$-differentiable function and increasing in $|e | > 0$.
\end{itemize} 
\end{definition} 

Let us first note that condition (L1) is equivalent with the statement
\beq \label{eq:rho}
\rho (x)=\rho_0(|x|) 
\eeq 
for some $\rho_0 : \R^+_0 \to  \R^+_0$.  The fact that  \eqref{eq:rho} $\Rightarrow$  (L1)  is obvious 
and the converse can be derived by invariance arguments.   
This illustrates  that $\rho$ is not $\C$-differentiable (i.e., holomorphic or analytic function). 
This is of course natural since 
only functions that are {\it both} holomorphic {\it and} real-valued are constants. 
The complex derivative of $\rho$ w.r.t. $x^*=(x_R + \im x_I)^*$ is  
 \[
 \psi(x) = \frac{\partial}{\partial x^*}\rho(x) = \frac 12 \left(\frac{\pr \rho}{\pr x_R} + \im  \frac{\pr \rho}{\pr x_I} \right)
 \]
 which will be referred in the sequel   as   the {\paino score function}.  Since $\rho(e)=\rho_0(|e|)$, 
 we can write $\psi$ using basic rules of complex differentiation \cite{eriksson_etal:2010}  in the form
\[
\psi(x)=  \frac 1 2 \rho_0'( |x|) \ssgn(x) ,
\]
where 
\[
\ssgn(e)= \begin{cases}   e/| e |, &\mbox{for  $e \neq  0$} \\     0 ,&\mbox{for $e = 0$} \end{cases} 
\]
is the  complex {\paino signum function}    and $\rho_0'$ denotes the real derivative of the real-valued function $\rho_0$.  
In order to make minimization of \eqref{eq:D}  possible by simple  gradient descent type algorithms, we narrow down the set of loss functions by imposing the assumption:
\begin{itemize}
\item[{\bf (L3)}] $\rho: \C \to \R^+_0$ is a convex function
\end{itemize} 
For example, the conventional LS loss function $\rho(x)=|x|^2$ verifies assumptions (L1)-(L3). In this case, $\rho_0(r)=r^2$ and  the score function is $\psi(x)= x$.
In this paper, we assume that the loss function verifies (L1)-(L3). 


We define {\paino Huber's loss function} in the complex case as 
\beq \label{eq:huber} 
\rho_{H,c}(e) =  
\begin{cases}  |e|^2, &\mbox{for  $|e| \leq c$} \\   2c |e| - c^2, &\mbox{for  $|e| > c$}, \end{cases}
\eeq 
where  $c$ is a  user-defined {\paino threshold} that influences the degree of robustness 
and efficiency of the method.  
Huber's function is a hybrid of $\ell_2$ and $\ell_1$ loss functions, using $\ell_2$-loss for relatively small errors and $\ell_1$-loss for relatively large errors. 
It verifies conditions (L1)-(L3). Huber's score ($\psi$-)function is  
\[
\psi_{H,c}(e) = \begin{cases} e, &\mbox{for  $|e| \leq c$} \\  c \,  \ssgn(e), &\mbox{for $|e|>c$}\end{cases}
\]
Note that Huber's $\psi$ is a winsorizing (clipping) funtion: the smaller the $c$, the more clipping is actioned on the residuals.

\section{Huber's criterion for multichannel sparse recovery} \label{sec:hub}

As discussed earlier, the scale $\sig$ of the error terms is unknown and needs to be estimated jointly with the signal matrix. We discuss here how this can be done elegantly using Huber's approach~2. 
First note that Maximum likelihood (ML-)approach for solving the unknown $\S$  and $\sigma$ leads to minimizing the negative log-likelihood function  of the form 
\[
 Q_{ML}(\S,\sig)  =   (\ndim \qdim) \log \sig + \sum_{i=1}^\ndim \sum_{j=1}^\qdim  \rho \! \left(  \frac{y_{ij} - \mb_{(i)}^\hop \s_i}{\sig} \right)  
\]
where $\rho(e)=-\log f_0(e)$ depends on the underlying standard form of the density $f_0(e)$ of the error terms. Then, one could replace the ML loss function $\rho$ with a robust loss function 
which need not be related to any circular density $f_0(\cdot)$, e.g., the Huber's loss function.  The negative log-likelihood function is however not convex in $(\S,\sig)$. This follows since $Q_{ML}(\S,\sig)$  is not convex in $\sig$ (for fixed $\S$)  and hence cannot be jointly convex. 
 
 Huber \cite{huber:1981} proposed an elegant devise to circumvent the above problem. See also \cite{owen2007robust} for further study of Huber's approach.  We generalize the Huber's approach 2 for the complex multivariate regression case and minimize 
\begin{align} \label{eq:Q}
Q(\S,\sig)  =   \al (\ndim \qdim) \sig  +  \sum_{i=1}^\ndim \sum_{j=1}^\qdim  \rho \! \left(  \frac{y_{ij} - \mb_{(i)}^\hop \s_i}{\sig} \right) \sig  ,
\end{align} 
 where $\al>0$Ê is a fixed {\paino scaling factor}.  Important feature of the objective function is that it  is 
 jointly convex in  $(\S,\sig)$ given that $\rho$ is convex.  In addition the minimizer $\hat \S$ preserves the same theoretical robustness properties (such as bounded influence function) as 
 the minimizer in the model where  $\sig$ is assumed to be known (fixed). 
 This is not the case for the ML-objective function $Q_{ML}(\S,\sig)$.

The stationary point of  \eqref{eq:Q} 
can be found by setting the complex matrix derivative of $Q$ w.r.t. $\X^*$  and 
the real derivative of $Q$ w.r.t. $\sig$ to zero. Simple calculations then show that the minimizer $(\hat \X,\hat \sig)$ is a solution to a pair of $M$-estimating equations: 
\begin{align}
\mm^\hop \psi\!\left( \frac{\E}{\sigma}\right) &= 
 \bo 0 \label{eq:estim1} \\  
\frac{1}{\ndim\qdim}\sum_{i=1}^\ndim \sum_{j=1}^\qdim \chi\! \left( \frac{y_{ij} - \mb_{(i)}^\hop \s_j}{\sig} \right) &= \al \label{eq:estim2} 
\end{align}
 where $\E =  \Y - \mm \X$ and   $\chi: \R_0^+ \to  \R_0^+$ is defined as
\beq \label{eq:chi}
\chi(t)=\rho_0'(t) t - \rho_0(t) . 
\eeq 
Recall that notation $\psi(\E)$ 
refers to element-wise application of  $\psi$-function to its matrix valued argument, so 
$[\psi(\E)]_{ij}=\psi(\ee_{ij})$.  
Thus if $\rho$ is convex and the MMV model is overdetermined with  non-sparse $\S$, solving the above $M$-estimating equations 
would give the global minimum of  \eqref{eq:Q}.

The scaling factor $\al$ in \eqref{eq:Q} is chosen so that the obtained scale estimate  $\hat \sig$ 
is Fisher-consistent for the unknown scale $\sigma$ when  $\eeps_{ij} \sim \C \mathcal N(0,\sig^2)$, which due to  \eqref{eq:estim2} 
is chosen so that 
\begin{align*}
\be &=  \expec[\chi(e)], \quad e\sim \C \mathcal N(0,1).  
\end{align*}
For many loss functions, $\be$ can be computed in closed-form.   
For example, for Huber's function \eqref{eq:huber}  the $\chi$-function in \eqref{eq:chi} becomes  
\[
\chi_{H,c}(e)=  |\psi_{H,c}(e) |^2 = \begin{cases}  |e|^2, &\mbox{for  $|e| \leq c$} \\    c^2, &\mbox{for  $|e| > c$}, \end{cases}
\]
and the concistency factor $\be=\be(c)$ can be easily solved in closed-form by elementary calculus as 
\begin{align} \label{eq:be} 
\be     &= c^2(1-F_{\chi^2_2}(2c^2)) + F_{\chi^2_4}(2c^2) .
\end{align}
Note that $\al$ depends on the threshold $c$. We will choose threshold $c$ as 
$
c^2 =  (1/2)F^{-1}_{\chi^2_2}(q)
$
for $q\in (0,1)$. The rationale behind this choice is that under Gaussian errors, $2 |e|^2/\sig^2 \sim  \chi^2_2$. Hence a sensible choice is to
 determine $c$ so that $2 c^2$ is the $q$th upper quantile 
of the $\chi^2_2$-distribution. The choice $q\to1$, implies $c^2 \to \infty$ and hence no-trimming of the residuals.   In our simulations we use $q=0.8$ which yields $c=1.269$. The smaller the $c$ (and hence $q$) the more trimming is actioned on residuals.

\section{SNIHT algorithm for Huber's criterion} \label{sec:SNIHT}

Our  aim is at solving 
\begin{align*}
\min_{\S, \sigma} Q(\S,\sigma)                 \mbox{ subject to}&  \ \| \S \|_0 \leq \kdim .   
\end{align*}
This problem is combinatorial (i.e., NP-hard) but {\paino greedy pursuit approaches} can be devised.  
Thus due to biconvexity of the objective function, we can use 
 Huber's loss function $\rho_{H,c}(e)$ 
 and greedy pursuit NIHT algorithm can be devised to compute an approximate solution. 
Recall that NIHT is a  {\paino projected gradient descent} 
method that is known to offer efficient and scalable solution for $\kdim$-sparse approximation problem \cite{blumensath_davies:2010}.  
NIHT updates the estimate of $\S$ by taking steps towards the direction of the negative gradient followed by projection 
onto the constrained space.  

In Huber's criterion, if we consider $\sig$ fixed at a value $\sig=\sig^{n+1}$ (the value of $\sig$ at $(n+1)$th iteration),   
the  simultaneous NIHT  (SNIHT) update of the signal matrix becomes
\[
\S^{n+1} = H_\kdim \big(\S^{n} \,  + \, \mu^{n+1} \mm^\hop \E_\psi^{n}\big)
\]
where $\mu^{n+1}$ is the update of the stepsize at $(n+1)$th iteration and 
\[
\E_\psi^{n} = 
 \psi \! \left( \frac{\E^n}{\sig^{n+1}} \right) \sig^{n+1} 
\]
will be referred to as {\paino pseudo-residual}. Note that 
$ -\nabla_{\X^*} \rho \! \left( \frac{\Y- \mm \S}{\sig^{n+1}} \right) (\sig^{n+1})^2 =  \mm^\hop\E_\psi^{n}$.  
The scale is updated (consider signal matrix $\X$ fixed at a value $\X=\X^n$) using \eqref{eq:estim2}  by a fixed-point iteration
\[
(\sig^{n+1})^2 = \dfrac{(\sigma^{n})^2}{\al} \dfrac{1}{ \ndim \qdim} {\displaystyle \sum_{i=1}^\ndim \sum_{j=1}^\qdim \chi \bigg( \frac{\ee_{ij}^{n}}{\sigma^n}  \bigg)}  ,
\]
where  $\E^{n} = \Y-\mm \S^{n}$  

The pseudo-code for the SNIHT algorithm in the case that the loss function $\rho$ is  Huber's function \eqref{eq:huber} is  given in Algorithm~\ref{Ollila:algor2}. We refer to this algorithm 
as  HUB-SNIHT in the sequel.  The steps 3-9 can be divided to 3 stages described below: 
{\paino scale stage} (Steps 3, 4) build up the  scale update $\sig^{n+1}$, 
{\paino signal stage} (Steps 5, 7, 8, 9) build up the $\kdim$-sparse signal update $\S^{n+1}$ and the support $\Gam^{n+1}$, and   
{\paino stepsize stage} (Step 7) computes the optimal stepsize update for the gradient descent move. The computation of the stepsize will be described in the next two paragraphs.   
Note that it is possible to tune the algorithm for different
applications by simply altering the criterion for
halting the algorithm. Matlab function is available at \url{http://users.spa.aalto.fi/esollila/software.html}. 


\begin{algorithm}
\caption{HUB-SNIHT algorithm} \label{Ollila:algor2}
\DontPrintSemicolon
\SetKwInOut{Input}{input}\SetKwInOut{Output}{output}
\SetKwFunction{Support}{InitSupport}
\SetKwInOut{Init}{initialize}
\SetAlgoNlRelativeSize{-1}
\SetNlSkip{0.4em}
\Input{ $\Y$,   $\mm$, sparsity $\kdim$, trimming threshold $c$. }
\Output{$( \S^{n+1}, \sig^{n+1}, \Gam^{n+1})$  estimates of $\S$, $\sig$ and $\Gam=\supp(\S)$.}
\Init{$\S^0=\bo 0$, $\mu^0=0$, $n=0$, $\Gam^0 = \emptyset$, 
$\al=\al(c)$.} 
 
 \BlankLine 
\nl $\sig^0 = 1.201 \cdot \mathrm{median}( |y_{ij}|, i=1,\ldots, \ndim, j=1,\ldots,\qdim)$

\nl $\Gam^0   = \rsupp(H_K \big( \mm^\hop \psi_{H,c}(\Y/\sig^0))\big)$     


\While{ halting criterion false}{\label{InRes1} 

\nl $\E^{n} = \Y-\mm \S^{n}$  

\nl $(\sig^{n+1})^2 = \dfrac{(\sigma^{n})^2}{\al} \dfrac{1}{ \ndim \qdim} {\displaystyle \sum_{i=1}^\ndim \sum_{j=1}^\qdim \bigg| \psi_{H,c} \bigg( \frac{\ee_{ij}^{n}}{\sigma^n}  \bigg)}  \bigg|^2 $

\nl $\E_\psi^{n} = \psi_{H,c} \bigg( \dfrac{\E^n}{\sig^{n+1}} \bigg) \sig^{n+1}$  

\nl $\G^n  = \mm^\hop \E_\psi^n$ 

\nl $\mu^{n+1}  =$ {\tt CompStepsize}($\E^n,\mm,\G, \Gam^n,\mu^n,\sig^{n+1}$)   

\nl $\S^{n+1} = H_\kdim (\S^{n} \,  + \, \mu^{n+1} \G^n )$  

\nl $\Gam^{n+1} = \rsupp(\S^{n+1})$ 

\nl $n=n+1$
 }
\end{algorithm}


As was noted in \cite{blumensath_davies:2010}, stepsize selection is very important for convergence and needs to be adaptively controlled at each iteration.  
Given the  found support $\Gam^n$  is correct,  we choose $\mu^{n+1}$  
as the minimizer of the convex objective function \eqref{eq:D} for fixed scale at $\sig^{n+1}$  
 in the gradient ascent direction $\S^n + \mu \G^n|_{ \Gamma^n}$, i.e. 
\begin{align} \label{eq:Lmu}
L(\mu)  &= D_{\rho_{H,c}}\!\! \left(\dfrac{\Y - \mm \left( \S^n + \mu\G^n|_{ \Gamma^n}  \right)}{\sig^{n+1}} \right) \notag\\
&= D_{\rho_{H,c}}\!\! \left(\dfrac{ \E^n - \mu \B^n}{\sig^{n+1}}  \right)  
\end{align} 
where $\E^n=\Y - \mm \S^n$ and  $\B^n = \mm_{\Gam^n} \G_{(\Gamma^n)}^n$. 
This reduces to minimizing  a simple linear  regression ($M$-)estimation problem   where  
the response is $\bo r=\vec(\E^n)$ and the  predictor is $\bo b=\vec(\B^n)$. 
It is easy to show (details omitted) that the minimizer $\hat \mu$ of $L(\mu)$ is the unique solution to a fixed point (FP) equation
$\mu = H(\mu)$, where 
\beq \label{eq:FP}
H(\mu) =  \left \| \B^n \right \|_{\W(\mu)}^{-2} \Re( \langle \E^n, \B^n \rangle_{\W(\mu)} ) 
\eeq 
where the right hand side  depends on  the unknown $\mu$ via the weight matrix $\W(\mu)$, defined as 
\[
\W(\mu) = w_{H,c} \bigg( \dfrac{\E^n - \mu \B^n}{\sig^{n+1}} \bigg) ,
\]
where $w_{H,c}$ is a weight function based on Huber's loss function, defined as
\[
w_{H,c}(e)  =\frac{\psi_{H,c}(e)}{e} =  \begin{cases} 1, &\mbox{for  $|e| \leq c$} \\  c/|e|, &\mbox{for $|e|>c$}\end{cases} .
\] 
If the loss function is LS-loss $\rho(e)=|e|^2$ (equivalent to Huber's function when $c\to \infty$), then 
the minimizer of \eqref{eq:Lmu} is easily found in closed form since in this case 
$\W(\mu)$ is equal to a matrix of ones.  Hence the FP equation is explicit and  the solution 
is $\mu^{n+1} =  \| \G_{(\Gam^n)}^n  \|^2 / \| \mm_{\Gam^n}   \G_{(\Gam^n)}^n \|^2$.  This  is  indeed 
the same stepsize used in  conventional SNIHT \cite{blanchard_etal:2014}.  

For Huber's loss function, the  minimizer of  \eqref{eq:Lmu} 
can be found 
by running the FP iterations  until convergence  (with  initial value $\mu_0>0$). Instead, we use approximate of the solution given by 
1-step FP iterate with initial value given by the previous stepsize $\mu^{n}$. In other words, 
in Step~7, the update $\mu^{n+1}$ is computed as 
$
\mu^{n+1}=H(\mu^n). 
$

\section{Application to Source Localization} \label{sec:array} 

We consider sensor array consisting of $\ndim$ sensors that receives $\kdim$ narrowband incoherent farfield plane-wave sources from a point source ($\ndim>\kdim$). 
At discrete time $t$,  the {\paino array output} (snapshot) $\y(t) \in \C^\ndim$ is a weighted linear 
combination of the signal waveforms $\s(t) = (\es_1(t), \ldots,\es_\kdim(t))^\top$ 
corrupted by additive noise $\eps(t) \in \C^\ndim$,  
$\y(t) = \A(\bom \theta) \s(t) + \eps(t)$,  
where  $\A=\bo A(\bom \theta)$ is the $\ndim \times \kdim$ {\paino steering matrix} para\-met\-rized by the vector $\bom \theta=(\theta_1,\ldots,\theta_\kdim)^\top$ 
of  (distinct) unknown direction-of-arrivals (DOA's)  of the sources. 
Each column vector $\a(\theta_i)$, called  the {\paino steering vector}, 
represents a point in known array manifold $\a(\theta)$.  
The objective of sensor array source localization is to find the DOA's of the sources, i.e.,  to identify the steering matrix $\A(\bth)$ 
parametrized by $\bth$. 
We assume  that the number of sources $\kdim$ is known.  
 
As in \cite{malioutov_etal:2005},   we cast the source localization problem as a multichannel sparse recovery problem. 
We construct an overcomplete $\ndim \times \pdim$ steering matrix $\A(\tilde \bth)$, where $\tilde \bth=(\tilde \theta_{1}, \ldots, \tilde \theta_{\pdim})^\top$ represents a sampling
grid of all source locations of interest. If $\tilde \bth$ contains the true DOA's $\theta_i$, $i=1,\ldots,\kdim$, then 
 the measurement matrix $\Y=\bmat \y(t_1) & \cdots & \y(t_\qdim) \emat \in \C^{\ndim \times \qdim}$ 
 consisting of snapshots at time instants $t_1,\ldots,t_\qdim$ can be {\it exactly} modelled as MMV model \eqref{eq:MMVmodel},   
where the signal matrix $\S \in \C^{\pdim \times \qdim}$ is $\kdim$-rowsparse matrix with  source signal sequences 
as its non-zero row vectors.
Thus identifying the source locations is equivalent to identifying the support $\Gam=\supp(\S)$ since any $i \in \Gam$ 
maps to a DOA $\tilde \theta_i$ in the grid.  Since the steering matrix  $\A(\tilde \bth)$ is completely known, we can use HUB-SNIHT method  to identify the support.

We assume that $\kdim=2$ independent (spatially and temporally) complex circular Gaussian source signals of equal power $\sigma^2_x$ arrive on an uniform linear array (ULA) of $\ndim=20$ sensors  with half a wavelength inter-element spacing from DOA's $\theta_1=0^o$ and $\theta_2=8^o$. 
In this case, the array manifold is $ \a(\theta) = ( 1,e^{-\im \pi  \sin(\theta)},\cdots,e^{-\im \pi (\ndim-1) \sin(\theta) })^\top$. 
The noise matrix $\Eps \in \C^{\ndim \times \qdim}$ has i.i.d. elements  following inverse Gaussian compound Gaussian (IG-CG) distribution 
\cite{ollila_etal:2012b} with shape parameter 
$\lambda=0.1$ and unit variance.   CG-IG distribution is heavy-tailed and  has been shown to accurately model radar clutter in  \cite{ollila_etal:2012b}. 
Note that the covariance matrix of the snapshot is $\cov(\y(t_i))= \sigma_x^2 \A(\bth)  \A(\bth)^\hop + \I_\ndim$, so we may 
use the popular MUSIC method to localize the sources. In other words, we search for  $\kdim=2$ peaks of the MUSIC pseudospectrum in the grid. 
We use a uniform grid $\tilde \bth$ on $[-90, 90]$ with 2$^o$ degree spacing, thus containing the true DOA's. For the source localization application, we make the following modifcation to the algorithm: In Step~1 of HUB-SNIHT algorithm, we locate the $K$ largest peaks 
of rownorms of $\mm^\hop\psi_{H,c}( \Y)$ instead of taking $\Gam^0$ as indices of $\kdim$ largest rownorms of $\mm^\hop\psi_{H,c}( \Y)$. 

\begin{figure}[!t]
\centerline{ \includegraphics[width=0.55\textwidth]{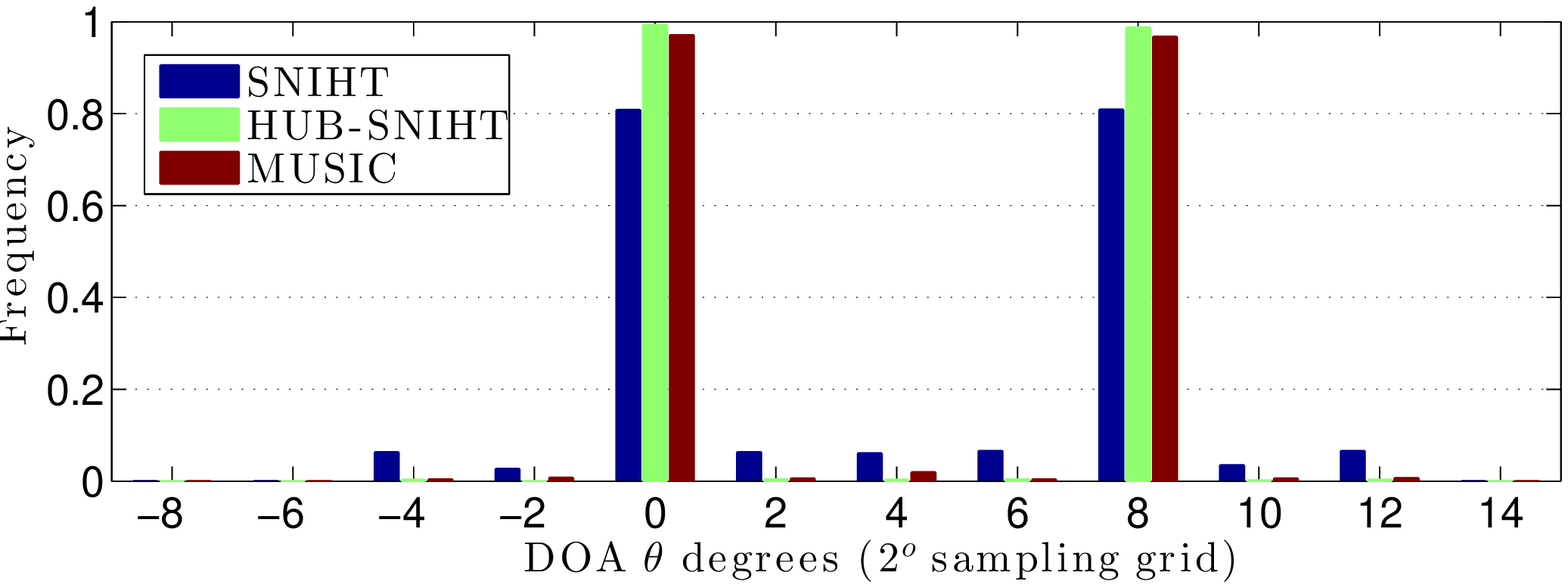}}
\centerline{ \includegraphics[width=0.55\textwidth]{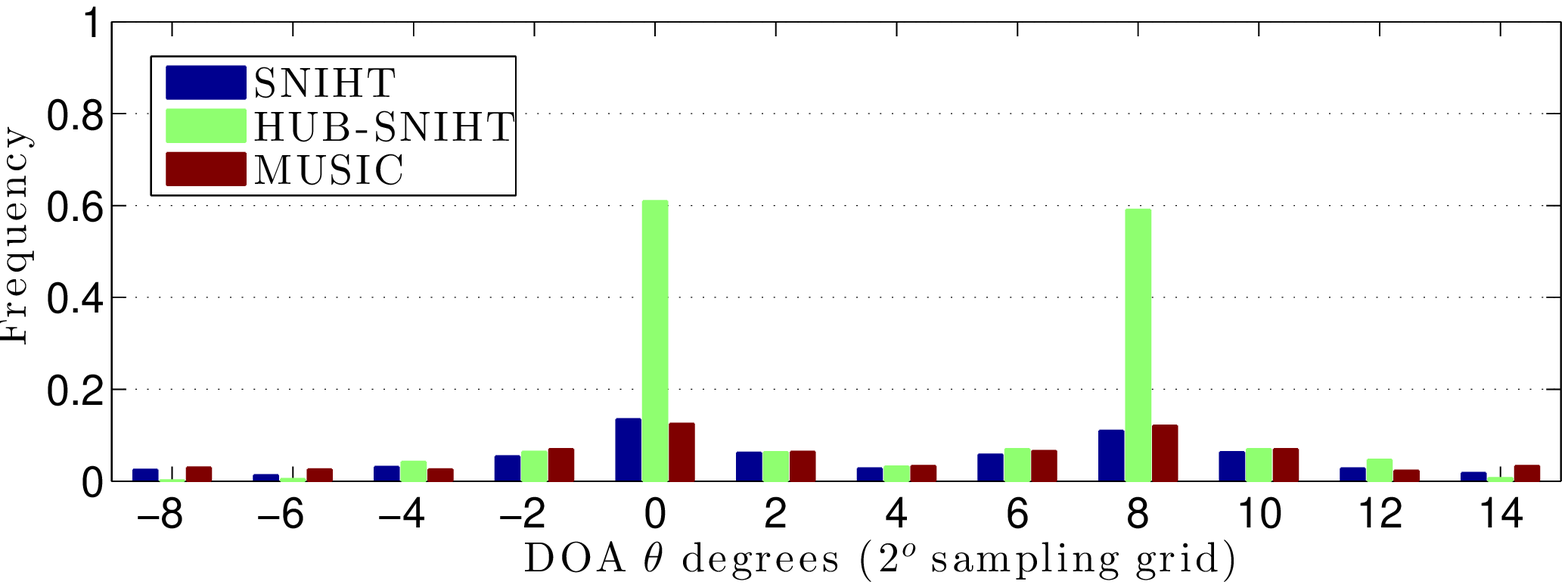}}
\centerline{ \includegraphics[width=0.55\textwidth]{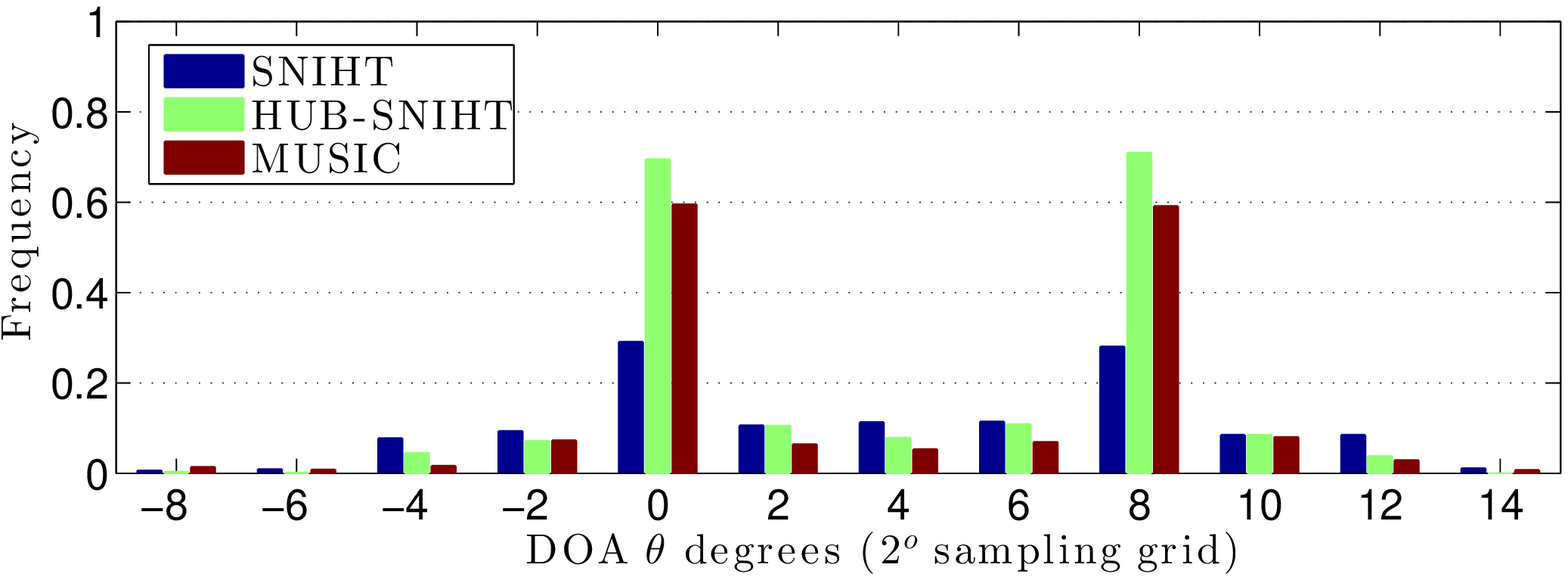}}
 \vspace*{-0.3cm}
\caption{Bar plots of relative frequency of DOA estimates. 
Two equal power  independent Gaussian sources arrive from  
 DOA  $0^o$ and $8^o$ and 
the noise has i.i.d. elements from  IG-CG distribution with unit variance and shape $\lambda=0.1$. 
 $\SNR =- 10$ dB and $\qdim=50$ (upper plot),   $\SNR =- 20$ dB and $\qdim=50$ (middle plot) and $\SNR =- 10$ dB and $\qdim=5$ (lower plot). 
}  \label{Ollila:fig3b}
\end{figure}

We then use SNIHT, HUB-SNIHT and MUSIC to identify the support (which gives the DOA estimates)  
and compute the empirical {\paino probability of exact recovery} (PER) rates and the relative frequency of DOA estimates in the grid 
based on 1000 MC runs. Full  PER rate $=1$  implies that the support $\Gam$  (and hence DOA's) were correctly identified  
in all MC trials. Such a case is shown in upper plot of Figure~\ref{Ollila:fig3b} for HUB-SNIHT when the number of snapshots is $\qdim=50$ and the SNR is $-10$ dB. 
The PER rate of HUB-SNIHT was $0.99$, but PER rates of SNIHT and MUSIC  were considerably lower, $0.81$ and $0.94$, respectively. 
In the second setting, we  lower the SNR to $ -20$ dB. 
In this case, the conventional SNIHT and MUSIC methods fail completely and provide nearly a uniform frequency on the grid. This is illustrated in the middle plot of Figure~1.
Note that the robust HUB-SNIHT provides high peaks on the correct DOA's. 
The PER rates  of SNIHT, HUB-SNIHT and MUSIC were  $0.02$,   $0.48$  and   $0.01$, respectively. Thus only HUB-SNIHT is able to 
offer good localization of the sources whereas the non-robust methods do not provide much better performance than a random guess. 
In the 3rd setting, we alter the set-up of 1st setting by decreasing the number of snapshots from $\qdim=50$ as low as $\qdim=5$. 
The performance differences between the methods are now more significant as is illustrated in the lower plot of Figure~\ref{Ollila:fig3b}. In this case the PER rates of SNIHT, HUB-SNIHT and MUSIC were  $0.19$,   $0.57$  and   $0.37$, respectively. Again, 
the HUB-SNIHT performed the best. 
 
  
\end{document}